# Ion-Pairing Limits Crystal Growth in Metal-Oxygen Batteries

Nagaphani B. Aetukuri[1,2] *, Gavin O. Jones[2], Leslie E. Thompson[2], Cagla Ozgit-Akgun[2,3], Esin Akca[2,3], Gökhan Demirci[2,3], Ho-Cheol Kim[2], Donald S Bethune[2], Kumar Virwani[2], Gregory M Wallraff[2]*

1. Solid State and Structural Chemistry Unit, Indian Institute of Science, Bengaluru, KA 560012 India
2. Advanced Energy Storage, IBM Research Almaden, San Jose, CA 95120 USA
3. ASELSAN Inc. – Microelectronics, Guidance and Electro-Optics Business Sector, Ankara 06750, Turkey

*Corresponding Author(s): *phani@iisc.ac.in* (N.B.A); *gmwall@us.ibm.com* (G.M.W)

*Aprotic alkali metal-oxygen batteries are widely considered to be promising high specific energy alternatives to Li-ion batteries. The growth and dissolution of alkali metal oxides such as $Li_2O_2$ in Li-$O_2$ batteries and $NaO_2$ and $KO_2$ in Na- and K-$O_2$ batteries, respectively, is central to the discharge and charge processes in these batteries. However, crystal growth and dissolution of the discharge products, especially in aprotic electrolytes, is poorly understood. In this work, we chose the growth of $NaO_2$ in Na-$O_2$ batteries as a model system and show that there is a strong correlation between the electrolyte salt concentration and the $NaO_2$ crystal size. With a combination of experiments and theory, we argue that the correlation is a direct manifestation of the strong cation-anion interactions leading to decreased crystal growth rate at high salt concentrations. Further, we propose and experimentally demonstrate that cation-coordinating crown molecules are suitable electrochemically stable electrolyte additives that weaken ion-pairing and enhance discharge capacities in metal-oxygen batteries while not negatively affecting their rechargeability.*

Over the past two decades, dramatic improvements to Li-ion batteries enabled the widespread adoption of mobile electronic devices and promise to revolutionize transportation. However, the high cost and limited specific energy of state-of-the-art Li-ion batteries are significant barriers for mass adoption as rechargeable energy storage devices for electrification of transportation[1]. Chemical conversion batteries such as aprotic metal-oxygen and metal-sulfur batteries are promising high specific energy alternatives to state-of-the-art Li-ion batteries[2-9]. Typically, in such chemical conversion batteries that offer high voltages, the discharge products could be electronically insulating due to the strong ionic interactions between the metal cation and the ligand anion. For example, in aprotic Li-$O_2$, Na-$O_2$ and K-$O_2$ batteries, the batteries' primary discharge products, $Li_2O_2$, $NaO_2$ and $KO_2$ respectively, are all electronically insulating[10-21]. The deposition of such electrically insulating discharge products could lead to cathode passivation and limit the maximum attainable discharge capacities[22]. While cathode passivation limited capacities have been experimentally observed in Li-$O_2$ batteries[23], Na-$O_2$ and K-$O_2$ batteries have shown discharge capacities much larger than those limited by cathode passivation[24,25]. In weakly-solvating aprotic electrolytes such as glymes, $Li_2O_2$ deposits as nanometer-scale thin conformal films in Li-$O_2$ batteries[26], while 3D growth of $NaO_2$ and $KO_2$ has been observed in Na-$O_2$ and K-$O_2$ cells[14,25]. Consequently, large discharge capacities are experimentally observed in Na-$O_2$ and K-$O_2$ batteries employing aprotic electrolytes.

A mechanistic understanding of the processes that enable electrochemical crystal growth in aprotic electrolytes could lead to the design of optimal electrolytes that afford large discharge capacities and also improve the rechargeability of all metal-oxygen batteries. However, the factors that influence electrolyte crystal growth in metal-oxygen batteries employing aprotic electrolytes are poorly understood. Perhaps, this is because electrolyte crystal growth in aprotic media, in general, is not widely studied. Fortunately, a large body of literature deals with the study of electrolyte crystal growth in aqueous solutions[27-29]. For example, the growth of AB crystals[30,31] such as NaCl or $CaCO_3$, with two components $A^+$ and $B^-$ in solution, has been extensively studied. However, there are at least two major differences between the growth of an electrolyte crystal from aqueous solutions and the crystal growth of discharge products from aprotic electrolyte solutions in metal-oxygen batteries. First, at low to moderate electrolyte concentrations, cation-anion pairing interactions can be neglected in aqueous solutions, while such ion-ion interactions are non-negligible in non-polar aprotic solvents. Second, electrolyte crystal growth is studied in electrolyte solutions with stoichiometric or near-stoichiometric ratios of the cations and anions. By contrast, in an aprotic metal-oxygen battery, the cation concentration ($M^+$) is determined by the electrolyte salt concentration while the anion ($O_2^-$) concentration is determined by its maximum solubility of ~1 mM in the aprotic solvent[32] (Note that in metal-oxygen batteries, $O_2^-$ is electrochemically generated by the reduction of molecular oxygen during discharge). In most cases, the electrolyte salt concentrations are >0.1 M and therefore the ratio of the alkali metal cations to superoxide anions is of the order of $10^2$ or larger.

Zhang and Nancollas[30] have developed analytical equations for the growth of AB crystals in such non-stoichiometric solutions. For the spiral growth of crystals in non-stoichiometric solutions, for example, it was shown that the linear growth rate, $R_g$, can be approximated by:

$$R_g = k_{in} \cdot \frac{a_B}{a_A} \cdot f(S) \quad (1)$$

where, $k_{in}$ is the integration controlled rate constant for crystal growth, $a_A$ and $a_B$ are the activities of the units $A^+$ and $B^-$ in solution with the restriction that $a_B < a_A$, and $f(S)$ is a saturation ratio ($S$) dependent function. For an AB crystal, the saturation ratio $S$ is given by: $\left(\frac{a_A \cdot a_B}{K_S}\right)^{1/2}$ where $K_S$ is the solubility product. $S$ is >1 for the growth of a crystal. The equation implies that the growth rates peak when the stoichiometry in solution is near unity. Depending on the integration rates of the individual growth units, the growth rate function need not peak near unity stoichiometry[33]. Nonetheless, for the purposes of this work it is important to note that, in general, increasing non-stoichiometry leads to a decrease in the crystal growth rate. This can be intuitively understood as follows: crystal growth requires alternate addition of the two growth units $A^+$ and $B^-$ on the surface of an existing crystal or a nucleus. The probability of integration of an ion on the surface of a crystal in solution can be approximated to be directly proportional to the ion's de-solvation frequency and the probability of its availability near a site of integration on the crystal surface such as a kink site. The probability of finding an ion near the site of integration will be smaller for the scarcer of the two ions and therefore the integration of the scarcer ion will be rate determining. In the specific case of crystal growth in metal-oxygen batteries, the integration of $O_2^-$ is the likely rate-determining step, as we discuss below.

During discharge in metal-oxygen batteries, in addition to the large difference in the concentrations of the $O_2^-$ and $M^+$ in solution, the presence of cation-anion pairing interactions could decrease the activity of superoxide anions in solution and further limit their availability for electrolyte crystal growth. For glymes with a dielectric constant (ε) of ~7.5 at room temperature (T=298 K), the critical radius for contact ion pair formation, which can be estimated from the Bjerrum theory of strong electrolytes[34,35], is ~37 Å (supplementary **section S2**). This implies that complete dissociation of ions in glymes is not possible if the shortest interionic distance of an ion-pair is less than 37 Å. The shortest interionic distances between the superoxide anion and any of the three alkali metal cations $K^+$, $Na^+$ and $Li^+$, as approximated from their ionic radii, are all about an order of magnitude less than this critical radius. Therefore, contact ion pairing will be a norm rather than an exception in glyme-based electrolyte solutions. By contrast, the critical radius is ~6 Å for a polar aprotic solvent such as dimethylsulfoxide (DMSO) (ε of ~47.2[36]). This is about the same order of magnitude as the shortest interionic distances between the superoxide anion and any of the three alkali metal cations $K^+$, $Na^+$ and $Li^+$. Therefore, contact ion pairing could be negligibly small in such solvents and solvent-separated or fully solvated ions the predominant species[34,37].

In this work, in order to investigate the effect of ion-ion interactions on crystal growth dynamics in metal-oxygen batteries, we have chosen the growth of $NaO_2$ crystals in $Na-O_2$ batteries employing glyme-based electrolytes as a model system. We chose $NaO_2$ as a model system because, i) the growth of $NaO_2$ can be directly interpreted within the framework of the growth of an AB crystal ($Na^+$ and $O_2^-$ being the growth units), ii) the formation of cuboidal $NaO_2$ is well known in $Na-O_2$ batteries and any changes in the size and/or morphology are accessible to characterization techniques such as scanning electron microscopy (SEM) and iii) experiments can be performed in batteries employing low dielectric constant (ε of ~7.5) glyme-based electrolytes where ion-ion interactions are expected to be the strongest. By employing electrolytes with different salt concentrations, we show that there is a strong correlation between the electrolyte salt concentration and $NaO_2$ crystal size during discharge. We propose that the weakening of ion pairing enables faster growth kinetics and enables $NaO_2$ crystal growth in $Na-O_2$ batteries employing electrolytes with lower salt concentrations. Furthermore, we show that suitable electrolyte additives that decrease ion-pairing tendency between $Na^+$ and $O_2^-$ increase the ultimate discharge capacities of $Na-O_2$ batteries. Our results suggest that additives or solvents that are electrochemically stable and also weaken ion pairing will improve discharge capacities in metal-oxygen batteries by enhancing crystal growth kinetics.

In order to probe the effect of ion-ion interactions on $NaO_2$ crystal growth, we have discharged $Na-O_2$ cells comprising a Na anode, a flat glassy carbon cathode (~1 cm$^2$) and electrolytes with concentrations of either 0.2 or 0.5 or 1 M sodium trifluoromethanesulfonate (sodium triflate or NaOTf) in 1,2 dimethoxyethane (DME) (As shown in supplementary **Fig. S1**, the cell impedance was found to increase with decreasing salt concentration and hence lower salt concentrations were not preferred). We have chosen glassy carbon electrodes over porous carbon electrodes to minimize current density variations across the electrode surface, thus allowing the changes in the sizes of the crystals to be directly correlated with the electrolyte concentration alone and not any local changes in current density or topography. The typical SEM images obtained on cathodes extracted from Na-O$_2$ cells after a galvanostatic discharge to 80 µAh/cm$^2$ at 20 µA/cm$^2$ are shown in **Fig. 1a-c**. Clearly, from the SEM images, it can be concluded that the NaO$_2$ particle size is the largest in the cathode extracted from the cell with the lowest electrolyte salt concentration. Simultaneously, the morphology also changes from growing along the directions perpendicular to the cube faces at the lowest electrolyte concentration to a more isotropic cube like morphology.

In order to understand the correlation between the electrolyte salt concentration and NaO$_2$ crystal size, we considered the following solution equilibria: 1) The equilibrium between the growing crystal surface and the growth units ($Na^+$ and $O_2^-$) in solution and 2) ion pairing interactions between the $Na^+$ and $O_2^-$ ions in solution. The two chemical equilibria are schematically shown in **Fig. 2** and can be represented as:

$(Na^+)_{sol} + (O_2^-)_{sol} \rightleftharpoons NaO_2(s)$ (2a)

$(Na^+)_{sol} + (O_2^-)_{sol} \rightleftharpoons (Na^+, O_2^-)_{sol}$ (2b)

The subscript *'sol'* represents a species in solution and *'s'* implies a solid phase. As shown in the schematic of **Fig. 2**, we assume that the growth of NaO$_2$ occurs by a sequential integration of $Na^+$ and $O_2^-$ on the growing NaO$_2$ crystal. Further, we neglect any contribution to the growth of NaO$_2$ by direct attachment of contact ion pairs and higher order aggregates as the diffusion constants of the ion-pairs and higher order aggregates are expected to be smaller due to their larger mass relative to the individual growth units. In the case of stoichiometric solutions (if the activities of $Na^+$ and $O_2^-$ are identical), the less mobile of the two ions controls the growth rate. However, the growth of NaO$_2$ occurs in a highly non-stoichiometric electrolyte solution containing nearly 2 orders of magnitude higher $Na^+$compared to $O_2^-$. In addition, as discussed earlier, the activity of $(O_2^-)_{sol}$ could be much less than the maximum solubility of $O_2^-$ in the electrolyte due to ion pairing with $Na^+$. Therefore the integration of $O_2^-$ is likely rate determining for NaO$_2$ crystal growth. Since the concentration of $(Na^+)_{sol} \gg$ concentration of $(O_2^-)_{sol}$ it is reasonable to assume that the activity of $(Na^+)_{sol}$ due to the ion-pairing interaction with $(O_2^-)_{sol}$ is nearly equal to the electrolyte salt concentration. And, it can be shown that the fraction of $O_2^-$ ($x_{(O_2^-)_{sol}}$), which is not participating in the ion pairing equilibrium and therefore available for the growth of $NaO_2(s)$ is given by[38] (also, see supplementary **section S3**):

$$x_{(O_2^-)_{sol}} = \frac{1}{1+K_A\cdot[(Na^+)_{sol}]} \quad (3)$$

where $K_A$ is the association constant for the ion pairing interaction (given by equation 2b) and is given by:

$$K_A = \frac{a_{(Na^+,O_2^-)_{sol}}}{a_{(O_2^-)_{sol}}\cdot a_{(Na^+)_{sol}}} \quad (4) \quad \text{where } a_i \text{ represents the activity of the species 'i'.}$$

Evidently, based on equation (3), the presence of ion pairing will decrease the fraction of $O_2^-$ available for the growth of $NaO_2$ crystals. It is worth emphasizing that the $x_{(O_2^-)_{sol}}$ will be significantly less than 1 for $K_A$ values >>1. Using Equations 1 and 3, the growth rate of $NaO_2$ in Na-$O_2$ batteries can be approximated by:

$$R_g = k_{in}\left(\frac{1}{[Na^+]\cdot[1+K_A\cdot[(Na^+)_{sol}]]}\right)f(S) \quad (5)$$

The order of magnitude of $K_A$, as estimated from the Bjerrum theory for strong electrolytes, is ~4.5 x $10^3$ for $Na^+$ and $O_2^-$ (supplementary **Table S1**). Using this value of $K_A$ in equation 3 shows that $x_{(O_2^-)_{sol}}$ can be as small as $10^{-3}$ and the growth rate as inferred from equation 5 will decrease by a similar order of magnitude ($10^{-3}$) in a 1 M $Na^+$ electrolyte solution. Furthermore, at such high values of $K_A$, $x_{(O_2^-)_{sol}}$ will also be sensitive to the salt concentration; a decrease in the salt concentration by an order of magnitude, for example, will increase the $x_{(O_2^-)_{sol}}$ by an identical order of magnitude. This would in turn increase the crystal growth rate (according to equation 5).

Therefore, for discharges performed in electrolytes with different salt concentrations, $NaO_2$ crystal sizes could be different: larger crystal sizes are expected at lower salt concentrations. This should be observable if 1) the discharges are performed at the same constant current density to the same ultimate capacity and 2) the concentration of the metal cations are still high enough such that the growth is not limited by the availability of a cation at the growth site. By using electrolyte concentrations >0.1 M and identical cathodes and discharge conditions, we have ensured that the experimental conditions closely match this theoretical picture. The SEM images, shown in **Figs. 1a-c**, are in agreement with this theory and clearly suggest that ion pairing could change crystal growth dynamics in $NaO_2$ batteries.

An alternate approach to weaken ion pairing would be to decrease $K_A$: decreasing $K_A$ should also enhance $NaO_2$ crystal growth. A suitable polar aprotic solvent such as DMSO or a solvent additive such as water will decrease ion pairing association constants ($K_A$) and likely enhance crystal growth. We have attempted to decrease ion-pairing association in $Na-O_2$ batteries, by adding trace quantities of a higher dielectric constant additive such as water to glyme-based electrolytes; this is an experiment similar to our work on understanding solvent effects in $Li-O_2$ batteries[26]. However, we observed that the trace water in electrolyte solvents reacted with the Na-anode, almost instantaneously, and made no observable difference to the size and morphology of $NaO_2$ or the rechargeability of $Na-O_2$ batteries[39]. Polar aprotic solvents such as dimethylsulfoxide (DMSO) and dimethylacetamide (DMAc) are reduced by Na-metal and therefore cannot be used as electrolyte solvents in $Na-O_2$ batteries, even if they might decrease ion pairing, increase superoxide solubility and possibly lead to large discharge capacities. Thus far, glyme solvents seem to be the only viable electrolyte solvents for $Na-O_2$ batteries.

We have identified crown ethers as suitable electrolyte additives to decrease ion-pairing association in glymes[40-43]. Crown ethers are chemically similar to glymes and are expected to be stable against sodium. In addition, crown ethers are known to coordinate alkali metal cations and could therefore weaken ion-pairing association when employed as electrolyte additives. Therefore, we have added 0.5 M of 1,4,7,10,13,16-hexaoxacyclooctadecane (18-Crown-6) as an electrolyte additive to a 1 M NaOTf in DME electrolyte and discharged a $Na-O_2$ cell with a glassy carbon electrode to a capacity of 80 µAh/cm$^2$ at a current density of 20 µA/cm$^2$. The SEM image of the cathode extracted from this cell is shown in **Fig. 1d**. Clearly, the size of the $Na-O_2$ cubes is larger than the $Na-O_2$ cubes obtained for a 1M NaOTf in DME electrolyte under identical galvanostatic discharge conditions. Furthermore, similar enhancement in the size of $NaO_2$ cubes was also observed on experiments performed with reduced graphene oxide (rGO) cathodes (supplementary **Fig. S2**). This is clear evidence that 18-Crown-6 (18C6) as an electrolyte additive enhances $NaO_2$ crystal growth.

In order to estimate the influence of 18C6 on decreasing the ion pairing association, we have performed computational calculations for the free energies of ion pairing of group I cations and 18C6-bound group I cations with superoxide in solution. These calculations involved the use of the unrestricted M06[44,45] density functional and geometries of all chemical species were optimized with the VTZ+[46] basis set and single point energies of optimized species were computed with the def2-tzvppd[46,47] basis set. All optimizations were performed in implicit tetrahydrofuran (THF) solvent (as a model for DME) with the SMD[48] method unless indicated otherwise.

The free energy for contact ion pair formation of group I cations with superoxide anion increases from $K^+$ to $Li^+$ in THF solvent as shown in **Table S2**. A similar trend is observed for the free energies for ion pairing in DMSO, but the pairing free energies are lower in DMSO. This is in agreement with our results shown in supplementary **Table S3**, which shows that the solvation energies of the alkali metal ions are larger in the more polar solvent suggesting that the electrostatic screening of the charge on the ion is better in the more polar solvent. Therefore, the driving force for ion pairing association is expected to be weaker in the more polar of the two solvents. These results are also fully consistent with the estimations based on Bjerrum theory shown in supplementary **Table S1** and strengthen the arguments that 1) ion-pairing association constants are non-negligible in glyme-based electrolytes and 2) ion pairing association constants for $Li^+$ and $O_2^-$ will be larger than for $Na^+$ and $O_2^-$. Further, we have calculated the free energy for the formation of a $O_2^-$-bound cation-crown complex when 18C6 is added to THF containing the ion pairs, $(M^+, O_2^-)_{sol}$. We found that the free energy for the formation of the complex is exergonic for the three group I cations that we studied (supplementary **Fig. S3**). This is further evidence that 18C6 as an additive weakens ion-pairing in glymes.

The trends observed for the ion pairing of $O_2^-$ with group I cations bound to 18C6 are also similar to ion-pairing with 'naked' group I cations: ion-pairing energy is still the largest for the $Li^+$ bound to 18C6 (**Fig. 3a**). This may be attributed to the trends observed earlier for 'naked' group I cations, but may also be attributed to the fact that $K^+$ is almost perfectly coordinated by the ring oxygen atoms of 18C6, while both $Na^+$ and $Li^+$ are asymmetrically bound (**Figs 3b-d**). The perfect coordination of $K^+$ with the ring oxygen atoms could render the ion less susceptible to being ion paired with $O_2^-$. Note that the free energies for ion-pairing between $O_2^-$ and the 18C6-bound cations shown in **Fig. 3a** are lower than those of the "naked cations". Ostensibly, this result is in agreement with our experimental observation that ion pairing is weakened in the presence of 18C6. Other crown ethers such as 1,4,7,10,13-pentaoxacyclopentadecane (15-Crown-5), which better coordinates $Na^+$ ions or cryptands such as 4,7,13,16,21,24-hexaoxa-1,10-diazabicyclo[8.8.8]hexacosane (cryptand C222), which have near 3D coordination could be much better at weakening ion association. However, our experiments with 15-Crown-5 or C222 as additives led to the precipitation of, presumably, salt-crown or -cryptand complexes and resulted in unstable electrolyte solutions. Therefore, we have continued to use 18C6 as a crystal growth-enhancing additive for the rest of the studies reported in this work.

In order to study the total capacity enhancement and electrochemical stability of 18C6 in $Na-O_2$ cells, we have assembled $Na-O_2$ cells comprising a Na anode, a macro-porous reduced graphene oxide (rGO) cathode and 1M NaOTf in DME electrolyte without or with 18C6 concentrations of 100 mM, 250 mM and 500 mM. In one set of experiments, the batteries were discharged to their ultimate capacities in oxygen ambient at 100 $\mu A/cm^2$ (see methods section for more details). The galvanostatic discharge curves obtained for various 18C6 concentrations are shown in **Fig. 4a**. The correlation between ultimate discharge capacities (normalized to the weight of rGO cathode) and the concentration of 18C6 in the electrolyte is plotted in the inset in **Fig. 4a**. Clearly, there is an increase in capacity by ~40% at the highest 18C6 concentration of 500 mM. SEM measurements (supplementary **Figs. S4 and S5**) on the cathodes after a discharge to full capacity showed that the discharge product completely covers the rGO cathode surface thereby blocking access to oxygen for further discharge reactions (so-called pore blocking limited capacity). Whether much higher capacity improvements are possible in optimized cathode structures is an open question.

In a second set of experiments, differential electrochemical mass spectrometry (DEMS) was performed during discharge and charge measurements for cells employing pristine 1M NaOTf in DME electrolyte and the 1M NaOTf in DME with 500 mM 18C6 as the additive. A comparison of the galvanostatic charge-discharge for the two cells is shown in **Fig. 4b**. The cells were discharged and charged to 1 mAh/$cm^2$ at a current density of 200 $\mu A/cm^2$. There is no discernable difference in the charge-discharge characteristics upon the addition of 18C6 to the electrolyte. Discharge $e^-/O_2$ plotted in **Fig. 4c** for the cells shows that the discharge is a 1$e^-/O_2$ process, which is clearly indicative of the formation of $NaO_2$[14,39,49]. In addition, both X-ray diffraction and Raman spectroscopy confirm that $NaO_2$ is the only observable discharge product[39] (supplementary **Figs. S6** and **S7**). There are no discernible differences in the spectroscopic signatures of the discharge product either structurally or chemically in the presence or absence of crown additives. In **Fig. 4c**, we also plot the summary of the DEMS measurements performed during the charge of Na-O$_2$ cells with the pristine electrolyte and the electrolyte with 0.5 M 18C6. Clearly, the rechargeability of the cell is unchanged by the addition of 18C6 as an additive. This is definite evidence that crown ethers are suitable electrochemically stable capacity-enhancing additives for Na-O$_2$ batteries.

The computational modeling showed that a similar weakening of ion pairing is also possible in Li-O$_2$ batteries employing 18C6 as an electrolyte additive. Therefore, we have performed galvanostatic discharge experiments in Li-O$_2$ cells comprising a Li anode, a macro-porous reduced graphene oxide (rGO) cathode and 1M LiOTf in DME without or with 18C6 in concentrations of 100 mM, 250 mM and 500 mM as the electrolyte. The relative capacity enhancement in Li-O$_2$ cells is even higher than the capacity enhancement observed for Na-O$_2$ batteries (supplementary **Fig. S8**). However, the highest ultimate gravimetric capacity obtained in Li-O$_2$ batteries is at least an order of magnitude lower than that obtained in Na-O$_2$ batteries prepared in this study. Furthermore, Li$_2$O$_2$ crystallite deposition has not been observed in SEM imaging (supplementary **Fig. S9**). Interestingly, the ion pairing association constant estimated from the Bjerrum theory for $Li^+$ and $O_2^-$ is an order of magnitude larger than that for $Na^+$ and $O_2^-$ (supplementary **Table S1**). Perhaps, the enhancement in capacity of Li-O$_2$ cells with the addition of 18C6 is strongly correlated with the ion-pairing association constants. However, we note that the disproportionation of LiO$_2$ to Li$_2$O$_2$ in Li-O$_2$ batteries complicates the study of discharge product growth and dissolution in these batteries.

The experimental and the theoretical data presented in this work provide clear evidence that disrupting ion pairing will, in general, enhance crystal growth in metal-oxygen batteries. This conclusion is also consistent with previous reports where it was found that lithium-oxygen batteries employing high donor number (DN) and/or high acceptor number (AN) solvents led to enhanced crystal growth[21,26,50]. Solvents with high DN (AN) better solvate metal cations ($O_2^-$ ions) thereby decreasing the ion-pairing strength. This would lead to an increased $x_{(O_2^-)_{sol}}$ that results in enhanced crystal growth as discussed earlier. In addition, three dimensionally coordinating additives such as C222 are expected to be more effective at coordinating metal ions and could further enhance crystal growth. However, there is a need to identify approaches to avoid salt precipitation in electrolytes with these 3D coordinating additives. It is also possible that such additives may enhance crystal dissolution and lead to improved rechargeability in metal-oxygen batteries.

**Acknowledgements**

The authors thank R. Miller (IBM Research) and J. Miller (Brookhaven National Lab) for discussions; I. Abate for Raman measurements; and the IBM model shop for support with the DEMS system. N.B.A. is supported by a faculty startup grant (No. 12-0205-0618-77) from the Indian Institute of Science.

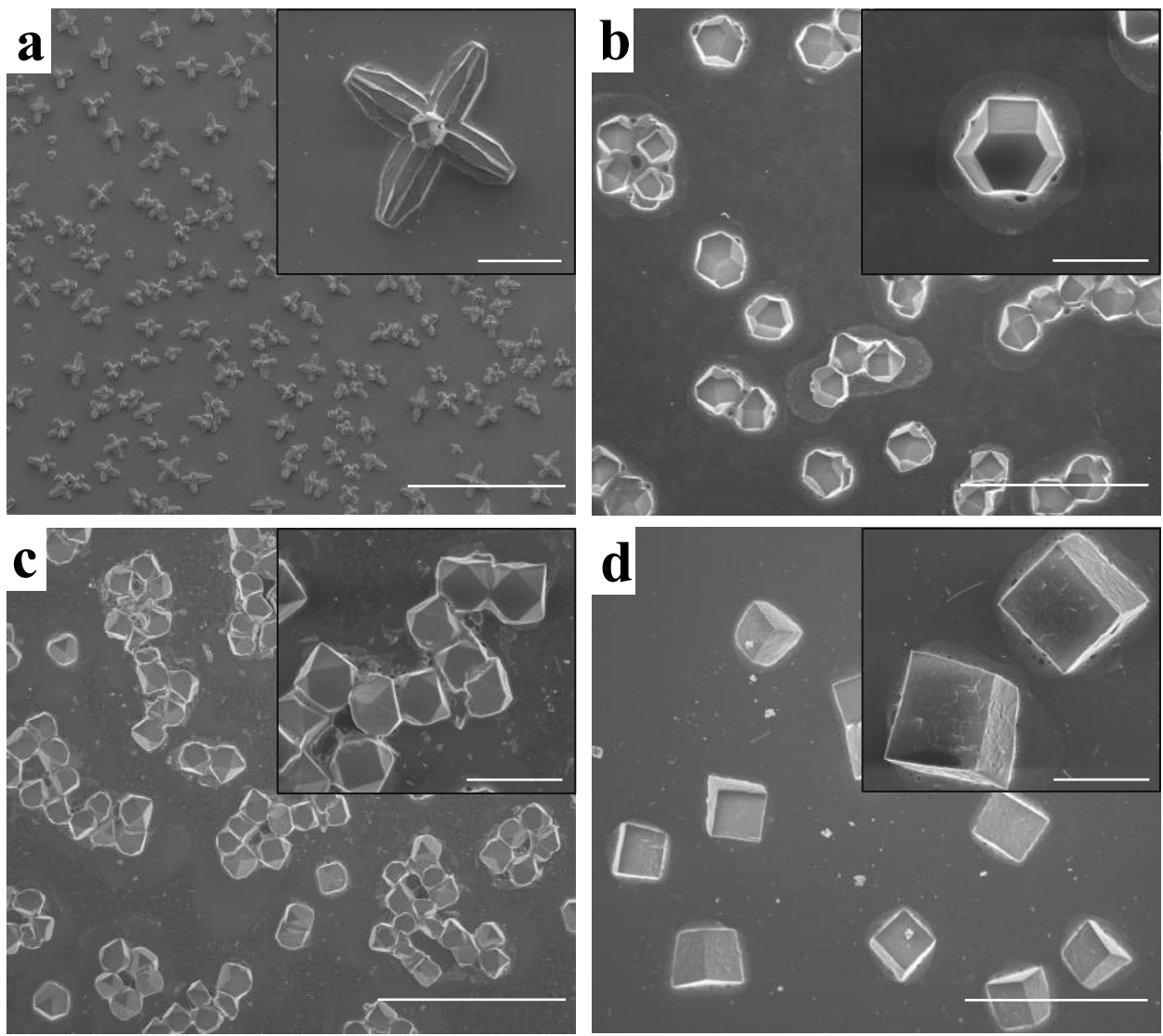

**Figure 1| NaO$_2$ growth control**. *a-d SEM images of NaO$_2$ deposits on glassy carbon electrodes extracted from Na-O$_2$ cells after a galvanostatic discharge to 80 µAh at 20 µA. The images correspond to cathodes extracted from cells employing (a) 0.2 M NaOTf in DME (b) 0.5 M NaOTf in DME (c) 1 M NaOTf in DME and (d) 1 M NaOTf in DME with 0.5 M 18C6 as an additive. The insets are higher magnification images of the respective electrodes. The scale bar for the SEM image in (a) is equivalent to 200 µm and for the image in inset of (a) is equivalent to 20 µm. The scale bars for SEM images in (b), (c) and (d) are all, equivalent to 40 µm and the scale bars for their insets are all equivalent to 10 µm. Since the NaO$_2$ crystal size is the largest for (a), images were acquired at lower magnification in order to image one complete NaO$_2$ crystal.*

**Figure. 1 Aetukuri *et al*.**

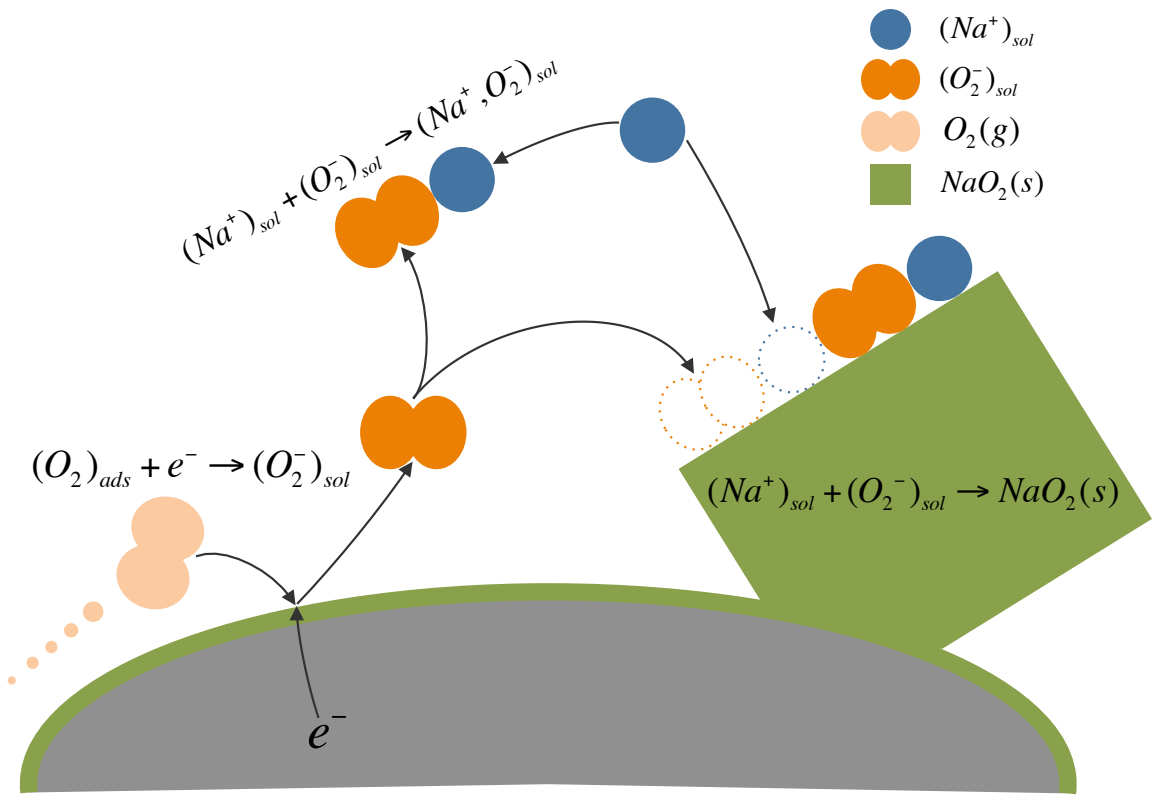

**Figure 2| Schematic showing NaO₂ solution equilibria near the cathode surface in aprotic Na-O₂ batteries**. *Oxygen adsorbed at an electrocatalytically active site on the cathode surface is reduced during the discharge step. The superoxide anion could either participate in an ion-pairing interaction with Na⁺ in the electrolyte solution or be available for the growth of NaO₂.*

**Figure. 2 Aetukuri *et al*.**

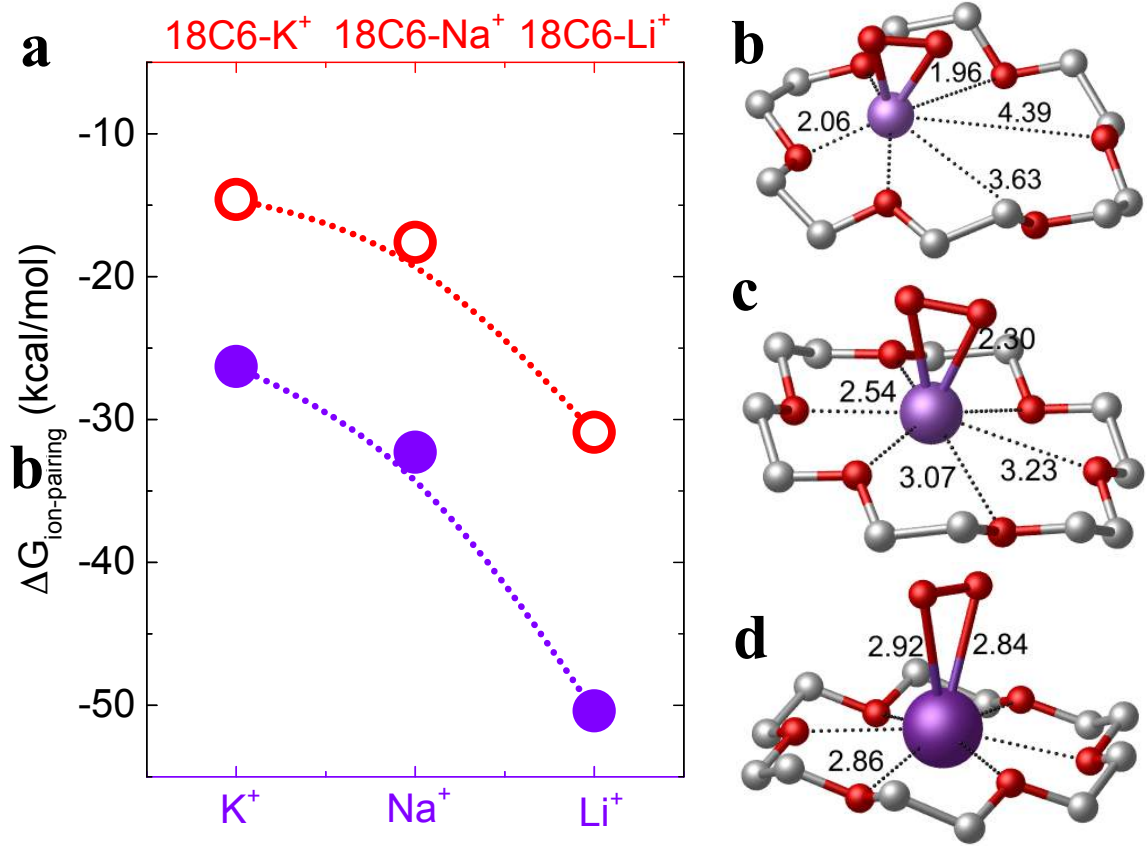

**Figure 3| Ion-pairing energies**. *a Plot of the free energy for ion-pairing for $K^+$, $Na^+$ and $Li^+$ with $O_2^-$ in THF with (red) and without (blue) the 18C6 coordinating the alkali metal cation. The ion-pairing free energy is less negative (suggesting weaker ion-pairing interaction) for the larger cations and for the cations coordinated by an 18C6 molecule. (b-d) Calculated structure of the cation-coordinated 18C6 molecule interacting with a superoxide anion clearly showing that (b) $Li^+$ and (c) $Na^+$ are not perfectly coordinated by the ring oxygen atoms of the 18C6 molecule while (d) $K^+$ is near-perfectly coordinated.*

**Figure. 3 Aetukuri *et al*.**

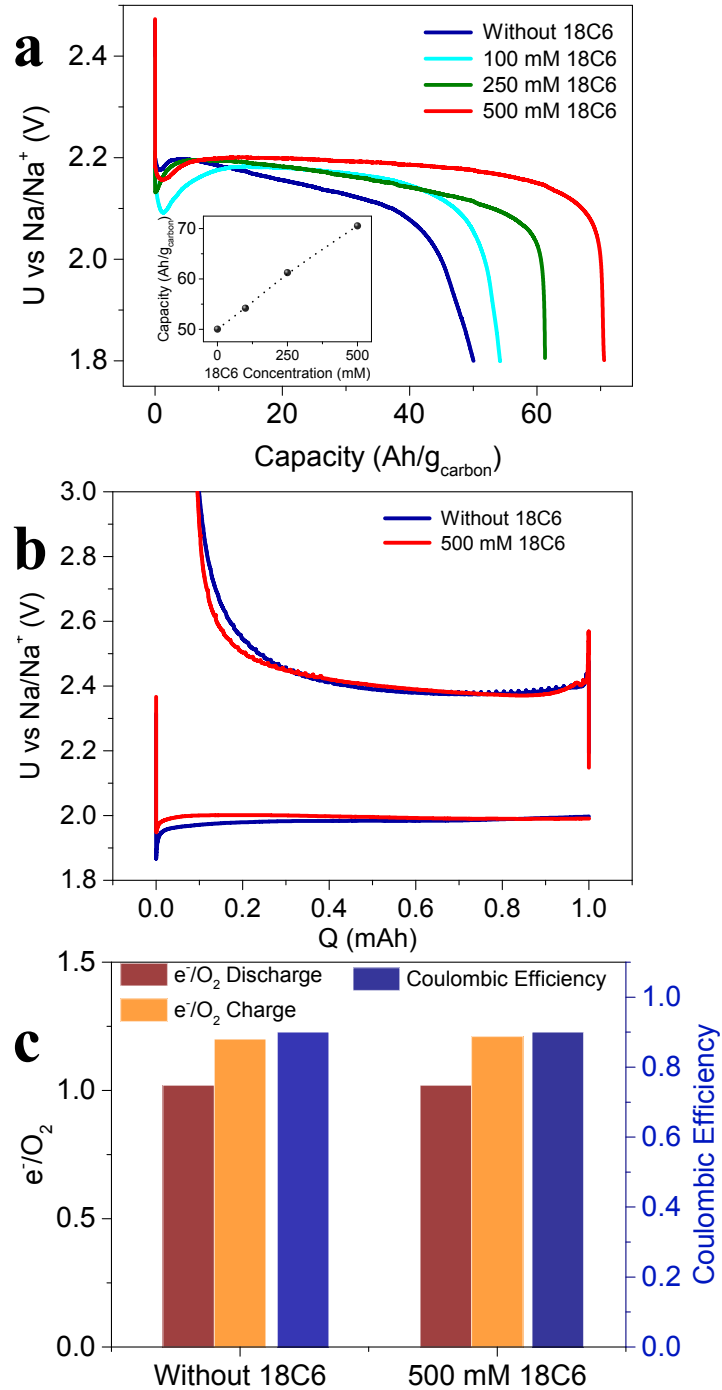

**Figure 4| Performance comparison of Na-O$_2$ batteries with and without 18C6 as an additive.** *a Discharge capacity at a current density of 100 µA/cm$^2$ for Na-O$_2$ cells with rGO cathodes and employing electrolytes without and with 100 mM, 250 mM and 500 mM of 18C6 as an electrolyte additive. The ultimate discharge capacities (normalized to the weight of rGO cathode) are plotted against the concentration of 18C6 in the inset in **Fig. 4a**. b Charge-discharge plots and c summary of DEMS analysis for Na-O$_2$ cells with rGO cathodes and employing 1M NaOTf in DME without and with 500 mM 18C6 as an electrolyte additive.*

**Figure. 4 Aetukuri *et al*.**

# Supplementary Information

# Ion-Pairing Limits Crystal Growth in Metal-Oxygen Batteries

Nagaphani B. Aetukuri[1,2] *, Gavin O. Jones[2], Leslie E. Thompson[2], Cagla Ozgit-Akgun[2,3], Esin Akca[2,3], Gökhan Demirci[2,3], Ho-Cheol Kim[2], Donald S Bethune[2], Kumar Virwani[2], Gregory M Wallraff[2]*

1. Solid State and Structural Chemistry Unit, Indian Institute of Science, Bengaluru, KA 560012 India
2. Advanced Energy Storage, IBM Research Almaden, San Jose, CA 95120 USA
3. ASELSAN Inc. – Microelectronics, Guidance and Electro-Optics Business Sector, Ankara 06750, Turkey

*Corresponding Author(s): *phani@iisc.ac.in* (N.B.A); *gmwall@us.ibm.com* (G.M.W)

## S1. Experimental Methods and Materials

Reduced Graphene Oxide Cathode Preparation

Reduced graphene oxide (rGO) cathodes were prepared by following the steps given below:

1. ~10 mg of single layer graphene oxide (GO) powder (ACS Material GnO1LP) was dispersed in 1 ml ultra-high purity MilliQ® water ($\rho = 18.2$ MΩ-cm) by sonication in an ultrasonic bath.
2. The solution was stirred overnight at room temperature (for approximately 12 hours) at 400 rpm using a magnetic stirrer.
3. Approximately 40-50 µl of GO solution were drop cast on to 12 mm in diameter stainless steel (316L grade) discs. The discs were dipped into liquid $N_2$ and then dried via freeze-drying in a Labconco Freeze Dry System, Freezone® 4.5 for 24 h in order to preserve the macroporous 3D structure of GO.
4. GO cathodes were then reduced at 600 °C for 3 h with Ar flow using a quartz tube furnace (Thermo Scientific, Lindberg Blue M).
5. The cathodes were transferred to the glove box while still warm (~60 °C) and were always kept on a hot plate at ~120 °C to avoid any solvent or trace moisture absorption prior to cell assembly.

Glassy Carbon Cathode Preparation

Double side polished 1 mm thick, 100 x 100 mm$^2$ glassy carbon plates were purchased from Tokai Carbon. 12 mm diameter discs were either laser or electric-discharge machined from these plates to enable their use as cathodes in our custom designed cells. Prior to each use in cells, glassy carbon discs were manually polished (in a figure 8 pattern) using alumina polishing suspension, as the abrasive agent, with a grit diameter of 1 μm (Buehler MicroPolish II 1 μm), and subsequently with alumina polishing suspensions of grit diameter 0.3 μm (Buehler MicroPolish II 0.3 μm) and finally with a suspension of grit diameter 0.05 μm (Buehler MasterPrep 0.05 μm). After polishing, the discs were ultrasonically cleaned in ultrapure deionized (DI) water (18.2 MΩ-cm) and then in a (50%/50%, v/v) mixture of ethanol and DI water. After sonication, glassy carbon surfaces were first blow dried with a N$_2$ gun and then transferred to a 120 °C hot air oven where they were dried for at least 12 hours before using for electrochemical measurements.

Electrolytes

Battery grade 1,2 dimethoxyethane (DME) was purchased from BASF Corporation. Before using as a solvent for electrolyte preparation, as received DME was dried for at least 48 hours over 3 Å molecular sieves (purchased from Sigma Aldrich) which were activated for at least 8 hrs at 500 °C. Sodium trifluoromethanesulfonate (NaOTf) 98% and lithium trifluoromethanesulfonate (LiOTf) 99.995% were purchased from Sigma Aldrich. The as received salts were dried at 95 °C in an Argon ambient (<0.1 ppm H$_2$O and <0.1 ppm O$_2$) for at least 24 hours. 1,4,7,10,13,16-hexaoxacyclooctadecane (18-Crown-6) was purchased from Sigma Aldrich and used as received. Electrolytes prepared using the dried DME and NaOTf and with or without 18-Crown-6 showed water contents <20 ppm as measured by a coulometric Karl-Fisher titration measurement (Metro Ohm).

Electrochemical Measurements

Electrochemical measurements were performed using a VMP3 BioLogic multi-channel potentiostat. In-house designed differential electrochemical mass spectrometry (DEMS) system was used for gas analysis. All cells were assembled in a glove box operating in Argon ambient with <0.1 ppm H$_2$O and <0.1 ppm O$_2$. ~65 μl of the electrolyte is used for all measurements. Sodium ingots (99.95% purity) were purchased from Sigma Aldrich. Small pieces of sodium were cut from the ingot and rolled into ~100-200 μm thick foils between two polypropylene sheets. 11 mm sodium anode chips were punched from the foils and used as anodes. For XRD and Raman measurements, AvCarb® P50 was used as the cathode. As received P50 cathodes were punched into 12 mm discs and ultrasonically cleaned in isopropyl alcohol. Cathodes were then dried for at least 24 hours in a vacuum oven at 130 °C. The cathodes were transferred to the glove box while still hot and were always kept on a hot plate to avoid any solvent or trace moisture absorption. Before using in Na-$O_2$ cells, Whatman GF/C grade glass microfiber filters, which were used as electrode separators, were treated similar to the P50 cathodes. Discharge measurements were performed in oxygen and charge in Argon, unless otherwise mentioned. Matheson purity grade oxygen and argon, purchased from Matheson Tri-Gas, were fed to the cells via in-line moisture traps (Matheson Pur-Gas, gas purity >6.0). A more detailed experimental procedure was reported previously[1].

X-ray Diffraction (XRD) Measurements

Na-$O_2$ cells employing P50 cathodes were discharged to 3 mAh/cm$^2$ at a discharge current density of 200 μA/cm$^2$. The cells were then quickly removed from the DEMS apparatus and sealed. They are then moved to an argon glove box where the cathodes were extracted and quickly rinsed in DME to wash off excess electrolyte salt. They are then loaded into sealed X-ray cells with a Kapton window for XRD measurements. XRD measurements were performed on a Bruker D8 discover X-ray diffractometer using graphite monochromated Cu-Kα X-rays. The X-ray beam spot of 650 μm in diameter was rastered over a sample area of ~2 x 2 mm$^2$ by oscillating the sample in the x-y plane during measurements.

Scanning Electron Microscopy

An FEI Helios Nanolab 400s system was used for SEM measurements. Glassy Carbon, rGO or P50 cathodes were extracted from discharged Na-$O_2$ cells in an Argon dry box. The cathodes were then rinsed in DME (~200 μL) and residual DME on the cathodes is evaporated by evacuating the vials containing the cathodes in a vacuum chamber connected to the glove box. The cathodes were loaded onto an SEM sample holder and transferred in a sealed container to the SEM laboratory. Then, the SEM sample holder with the cathodes is loaded into the SEM sample-loading chamber and the latter is immediately evacuated. The total time for the transfer of the cathodes from the sealed container to sample loading chamber is minimized (usually <5 seconds). For samples exposed to ambient for much longer than this, the characteristic Na$O_2$ cube morphology was not observed.

## S2. Ion-Pairing Association Constant Calculation Within Bjerrum Theory for Glymes

*Calculation of critical ion-pair radius for complete dissociation in glymes*

According to Bjerrum theory[2,3], the critical radius ($r_c$) of an ion-pair for complete dissociation of ions in a solvent is given by:

$$r_c = \frac{q^2}{8\pi \cdot \varepsilon_0 \cdot \varepsilon \cdot k \cdot T} \qquad (S1)$$

For glymes with a dielectric constant ($\varepsilon$) of ~7.5 at room temperature (T=298 K), the critical radius is ~ 37 Å. This implies that complete dissociation of ions in the glymes is not possible if the shortest interionic distance of an ion-pair is less than 37 Å. The shortest distance between $Na^+$ and $O_2^-$ and $Li^+$ and $O_2^-$ is about an order of magnitude less than this critical radius and therefore ion-pairing will be a norm rather than exception in glyme solvents. By contrast, the critical radius is ~6 Å in dimethyl sulfoxide ($\varepsilon$ of ~47.2) which is about the same order as the shortest interionic radius of $Na^+$ and $O_2^-$ and $Li^+$ and $O_2^-$.

*Calculation of ion pairing association constants for alkali metal superoxides in glymes*

Following Fuoss and Krauss[3], we approximate the association constant ($K_A$) for ion-pairing by:

$$K_A = \left(\frac{4\pi \cdot N \cdot a^3}{1000 \cdot b}\right) \cdot exp(b) \qquad (S2)$$

where,

$$b = \frac{q^2}{4\pi \cdot \varepsilon_0 \cdot \varepsilon \cdot a \cdot k \cdot T} \qquad (S3)$$

Ion-pairing association constants were calculated for the different alkali metal ions and superoxide anion. The radius of the superoxide anion is taken as[4]: 1.71 Å (see supplementary **Table. S1**)

*Definitions of symbols used in the above equations S1-S3.:*

$N$: Avogadro's number

$\varepsilon_0$: Permittivity of free space

$\varepsilon$: Dielectric constant of the solvent

$q$: Elementary charge

$k$: Boltzmann Constant

$a$: Shortest distance between the ion centers of an ion-pair

$T$: Temperature

## S3. Relationship Between the Activity of $O_2^-$, the Ion-Pairing Constant and the Metal-Ion Concentration

The ion-pairing association constant for the chemical equilibrium: $(O_2^-)_{sol} + (M^+)_{sol} \leftrightarrow (M^+, O_2^-)_{sol}$ is given by:

$$K_A = \frac{a_{(M^+,O_2^-)_{sol}}}{a_{(O_2^-)_{sol}} \cdot a_{(M^+)_{sol}}} \qquad (S4)$$

where $(O_2^-)_{sol}$, $(M^+)_{sol}$ and $(M^+, O_2^-)_{sol}$ are the superoxide anion, the alkali metal cation and the ion pair in solution and $a_i$ represents the activity of the species '$i$'. Assuming that higher order ion pair formation is negligible, ion and ion-pair activities can be approximated by their respective mole fractions in solution[5]. The fraction of soluble $O_2^-$ which is not interacting with $M^+$ in the ion-pairing equilibrium is given by (the terms in the square brackets represent concentration of the respective species):

$$x_{(O_2^-)_{sol}} = \frac{[(O_2^-)_{sol}]}{[(O_2^-)_{sol}] + [(M^+, O_2^-)_{sol}]} \qquad (S5)$$

Combining equations S4 and S5,

$$x_{(O_2^-)_{sol}} = \frac{[(O_2^-)_{sol}]}{[(O_2^-)_{sol}] + K_A \cdot [(O_2^-)_{sol}] \cdot [(M^+)_{sol}]} \qquad (S6)$$

Simplifying equation S6 gives:

$$x_{(O_2^-)_{sol}} = \frac{1}{1 + K_A \cdot [(M^+)_{sol}]} \qquad (S7)$$

Therefore, in the presence of ion-pairing, the activity of the free ions of $O_2^-$ in solution (i.e., the $O_2^-$ available for the growth of oxide discharge product) is decreased by a factor of $1 + K_A \cdot [(M^+)_{sol}]$.

## S4. Supplementary Figures

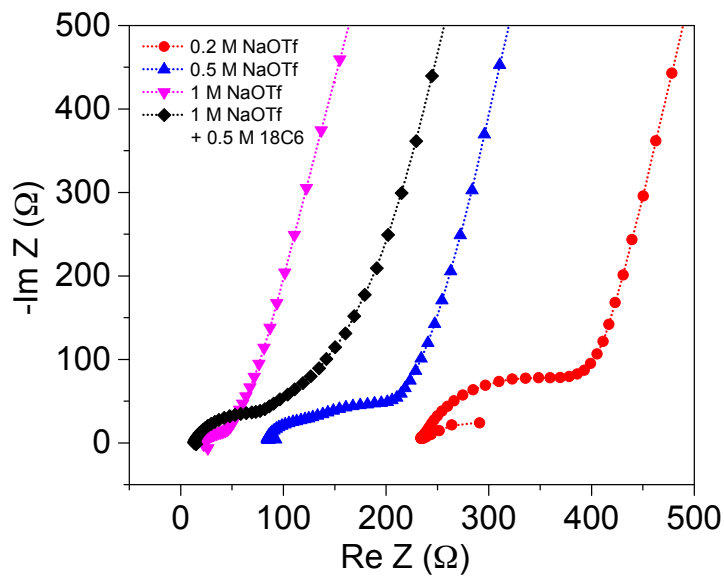

**Figure S1|** *Impedance spectra of of Na-O$_2$ cells with glassy carbon cathodes and employing electrolytes of various salt concentrations and with or without 500 mM of 18C6 as shown in the figure's legend. Clearly, the impedance is much higher for electrolytes with low salt concentrations. The addition of 0.5 M 18C6 does not significantly alter the impedance.*

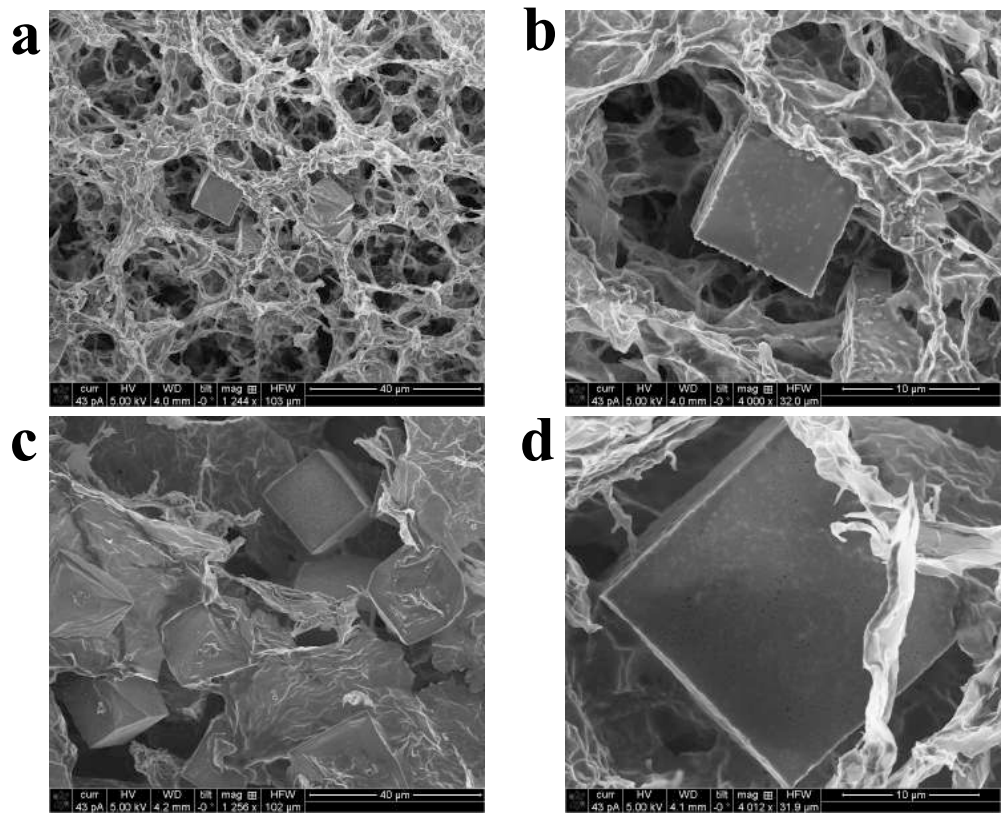

**Figure S2|** *SEM images of $NaO_2$ deposits on rGO cathodes extracted from $Na$-$O_2$ cells after a galvanostatic discharge to 200 µAh/cm$^2$ at 10 µA/cm$^2$. The images correspond to cathodes extracted from cells employing **a, b** 1 M NaOTf in DME and **c, d** 1 M NaOTf in DME with 0.5 M 18C6. Clearly, crystallite size increases upon the addition of 18C6 as an electrolyte additive.*

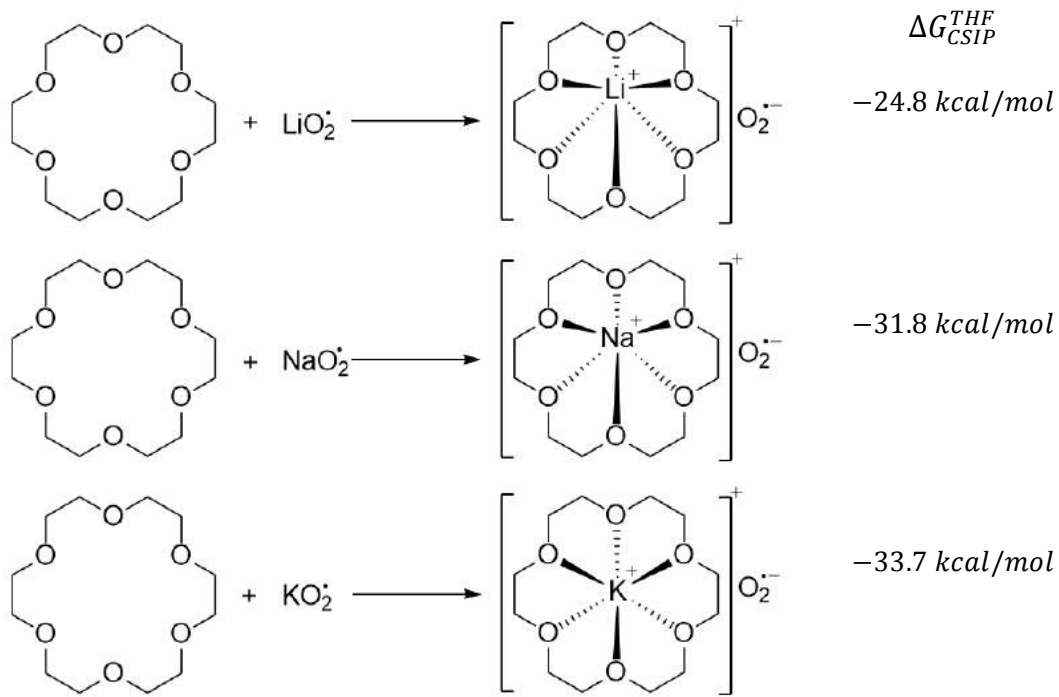

**Figure S3|** *Chemical reaction schemes that lead to 18C6 (crown) separated ion-pair (CSIP) formation in the presence of 18C6. The calculated free energies for these reaction schemes are all negative suggesting that the addition of 18C6 to an electrolyte containing alkali metal-superoxide ion pairs will spontaneously lead to crown-separated ion-pair formation.*

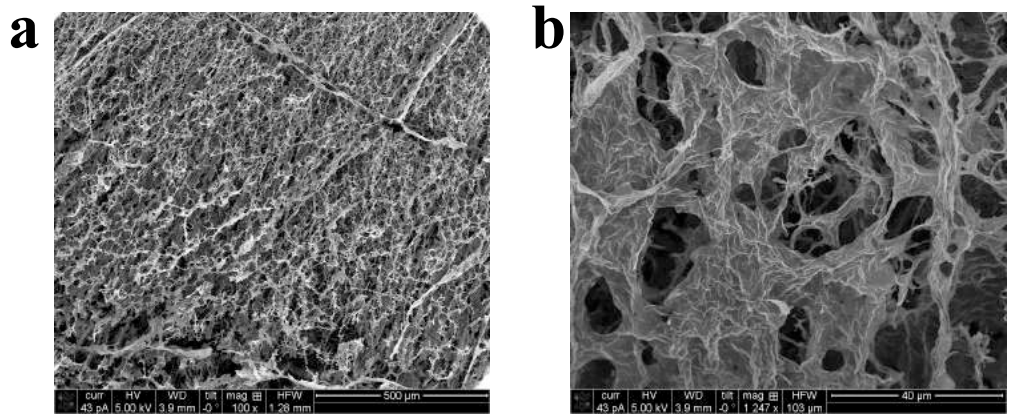

**Figure S4|** *SEM images of as prepared rGO cathodes at two different magnifications. The highly macro porous nature of the cathodes can be seen.*

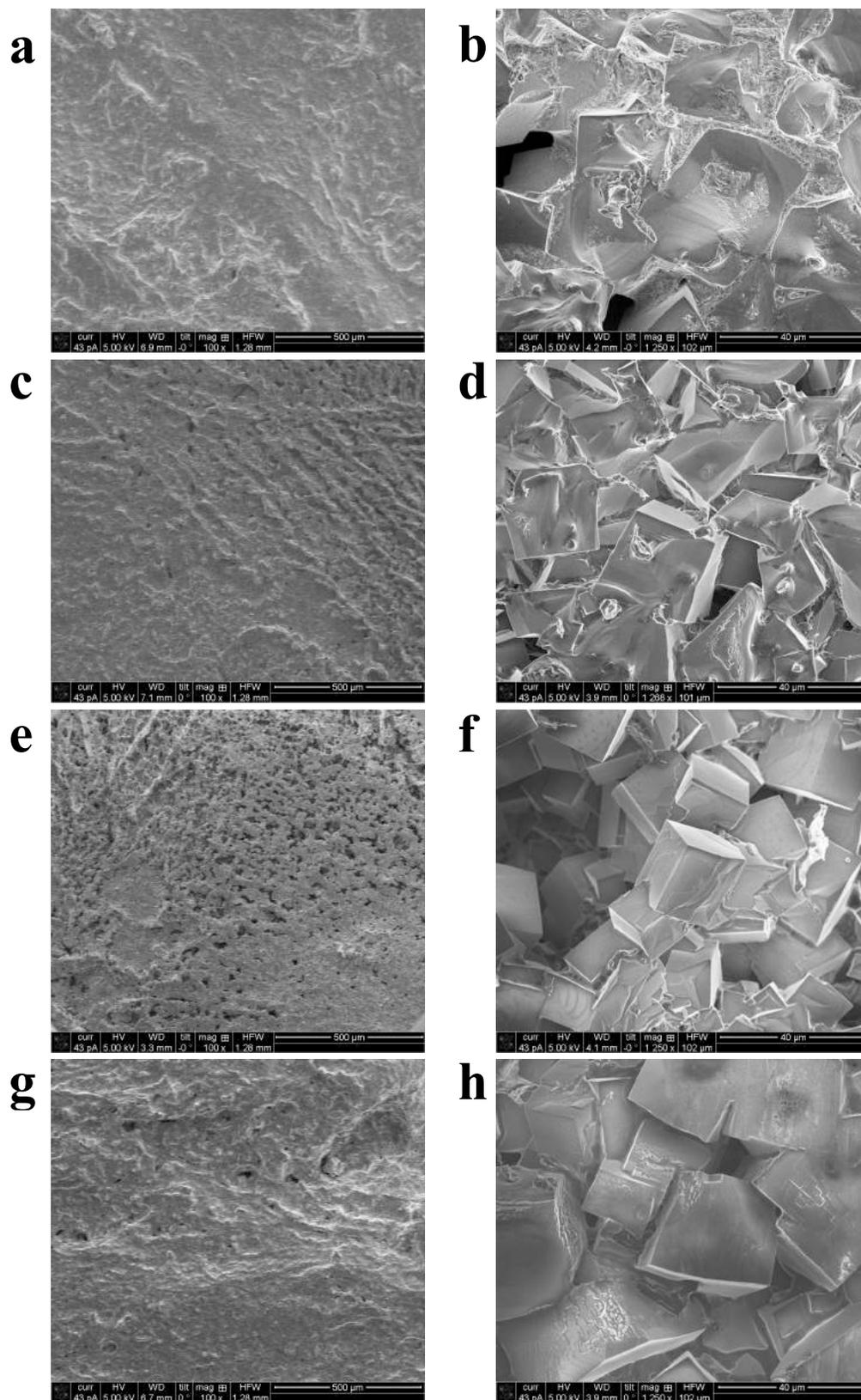

**Figure S5|** *SEM images of rGO cathodes extracted from Na-O₂ cells after a galvanostatic discharge at 100 μA/cm² to ultimate discharge capacity. The images correspond to cathodes extracted from cells*

employing *a, b* 1 M NaOTf in DME; *c, d* 1 M NaOTf in DME with 0.1 M 18C6; *e, f* 1 M NaOTf in DME with 0.25 M 18C6; *g, h* 1 M NaOTf in DME with 0.5 M 18C6. In all cases, it seems that pore blocking has limited the ultimate discharge capacities.

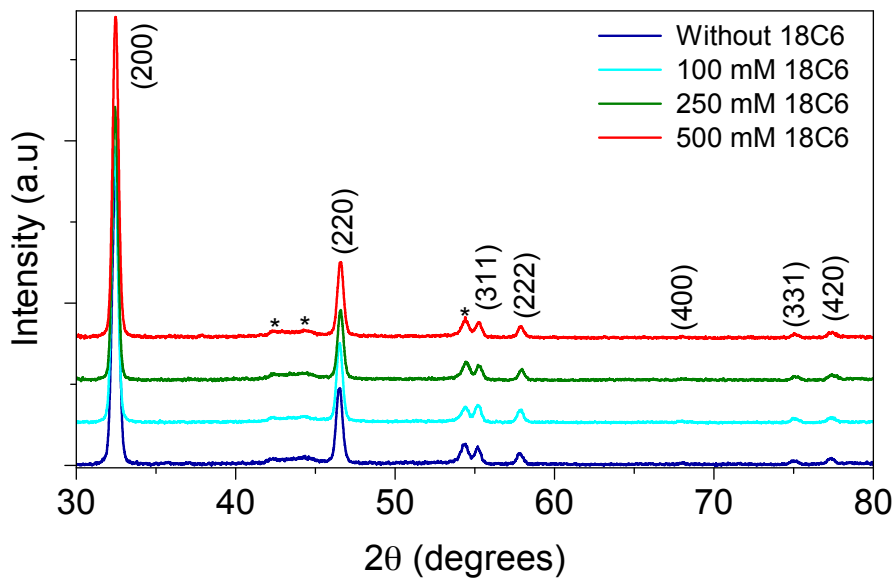

**Figure S6|** *X-ray diffractograms of cathodes extracted from Na-$O_2$ cells employing 1 M NaOTf in DME electrolytes with different concentrations of 18C6 as an electrolyte additive. All cells were discharged to 3 mAh at 200 µA/$cm^2$. All the XRD peaks can be indexed to the distorted pyrite-type structure of $NaO_2$ (JCPDS 01-077-0207). The peaks identified with asterisks (\*) are from the P50 carbon cathode.*

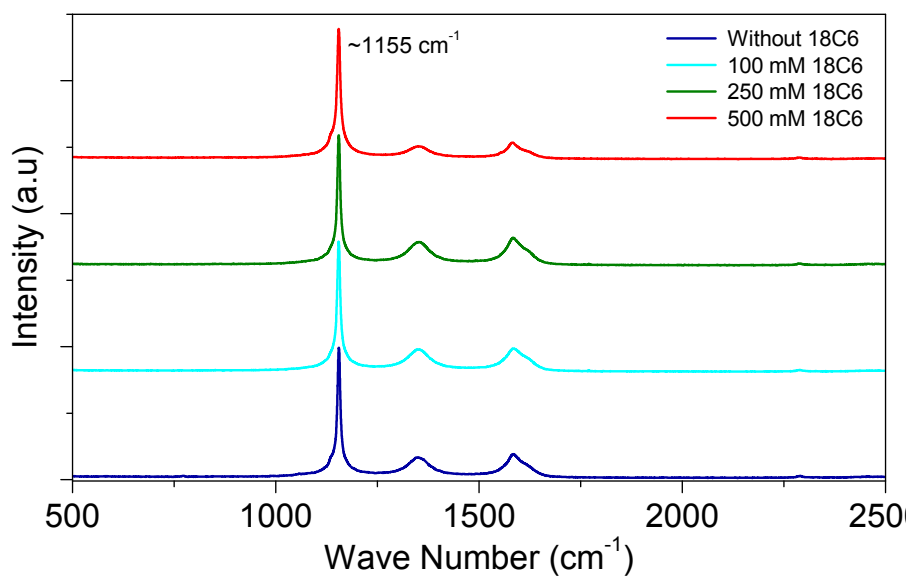

**Figure S7|** *Raman spectra of cathodes extracted from Na-O$_2$ cells employing 1 M NaOTf in DME electrolytes with different concentrations of 18C6 as an electrolyte additive. All cells were discharged to 3 mAh at 200 µA/cm$^2$. The Raman peak at ~1155 cm$^{-1}$ corresponds to NaO$_2$. The two peaks at higher wave numbers are from the P50 carbon cathode.*

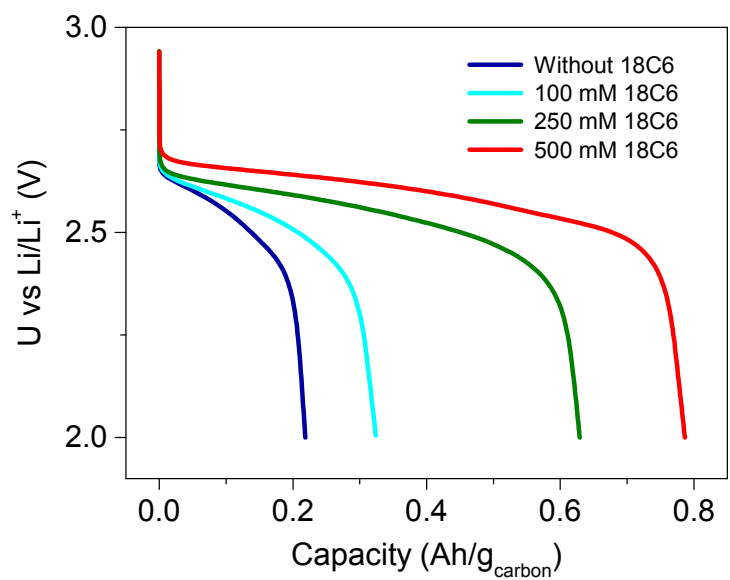

**Figure S8|** *Discharge capacity of Li-O$_2$ cells with rGO cathodes and employing electrolytes without and with 100 mM, 250 mM and 500 mM of 18C6 as an electrolyte additive. Clearly, addition of 18C6 enhances ultimate discharge capacities in Li-O$_2$ cells.*

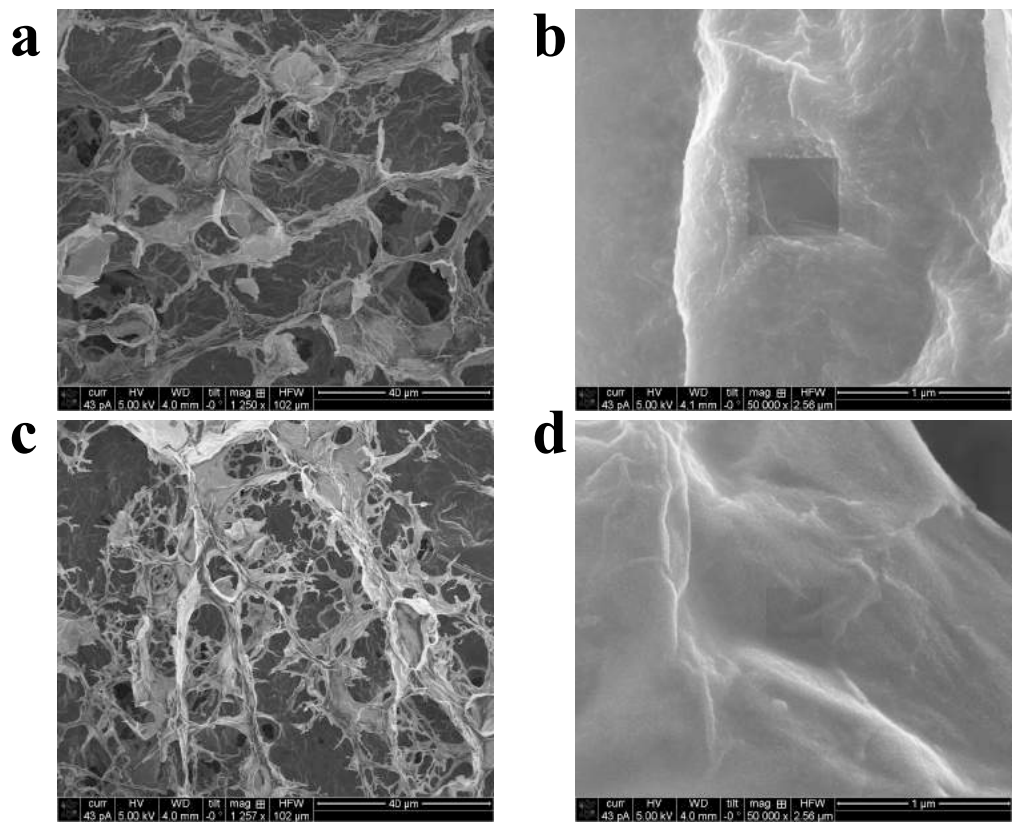

**Figure S9|** *SEM images of rGO cathodes extracted from Li-$O_2$ cells after a galvanostatic discharge to 200 µAh/$cm^2$ at 10 µA/$cm^2$. The images correspond to cathodes extracted from cells employing **a, b** 1 M LiOTf in DME and **c, d** 1 M LiOTf in DME with 0.5 M 18C6. There is deposition of $Li_2O_2$ during discharge as evidenced from the contrast difference between most of the cathode surface and a small square (in **b, d**) where the $Li_2O_2$ is deliberately decomposed by the SEM electron beam. Crystallites of $Li_2O_2$ have not been observed in SEM images.*

## S5. Supplementary Tables

| Alkali Metal Ion | Ionic Radius (Å) | inter-ionic distance (Å) | b | $K_A$ ($mole^{-1}$) |
|---|---|---|---|---|
| $Li^+$ (4-fold Coordinated) | 0.59 | 2.3 | 32.53 | 3.8 x $10^5$ |
| $Li^+$ (6-fold Coordinated) | 0.76 | 2.47 | 30.29 | 5.4 x $10^4$ |
| $Na^+$ (6-fold Coordinated) | 1.02 | 2.73 | 27.41 | 4.5 x $10^3$ |
| $K^+$ (6-fold Coordinated) | 1.38 | 3.09 | 24.21 | 3.0 x $10^2$ |

**Table S1|** *Calculations of association constants for ion pairing between the superoxide anion and different alkali metal ions as listed in the table. The sizes of transition metal ions have been taken from*

*Shannon 1976[6]. A six-fold coordination of alkali metal ions by the superoxide anion is considered, except for $Li^+$ where both 4-fold and 6-fold coordinations were considered.*

| Cation | $\Delta G_{CIP}^{THF}$ (kcal/mol) | $\Delta G_{CIP}^{DMSO}$ (kcal/mol) |
|---|---|---|
| $K^+$ | -26.3 | -13.6 |
| $Na^+$ | -32.3 | -18.3 |
| $Li^+$ | -50.4 | -34.1 |

**Table S2|** *Table showing calculated ion-pairing free energies for contact ion-pair (CIP) formation in THF and DMSO for the forward reaction: $M^+(sol) + O_2^-(sol) \rightarrow [M^+, O_2^-](sol)$ where $M^+$ is an alkali metal cation. The driving force for ion-pair formation is lower in DMSO when compared to THF for all the alkali metals considered for these calculations.*

| Alkali Metal Ion | $\Delta G_{solvation}^{THF}$ (kcal/mol) | $\Delta G_{solvation}^{DMSO}$ (kcal/mol) |
|---|---|---|
| $Li^+$ | -66.0 | -74.3 |
| $Na^+$ | -52.4 | -58.7 |
| $K^+$ | -42.5 | -47.4 |

**Table S3|** *Calculated solvation energies of various alkali metal ions in THF and DMSO as referenced against a free ion in vacuum.*